# Terabit Indoor Laser-Based Wireless Communications: LiFi 2.0 for 6G

[1]Mohammad Dehghani Soltani, [2]Hossein Kazemi, [2]Elham Sarbazi, [3]Ahmad Adnan Qidan,
[3]Barzan Yosuf, [3]Sanaa Mohamed, [4,6]Ravinder Singh, [7]Bela Berde, [7]Dominique Chiaroni,
[8]Bastien Béchadergue, [9]Fathi Abdeldayem, [10]Hardik Soni, [11]Jose Tabu, [12]Micheline Perrufel,
[13]Nikola Serafimovski, [3]Taisir E. H. El-Gorashi, [3]Jaafar Elmirghani, [4]Richard Penty, [4,5]Ian H. White,
[2]Harald Haas and [1]Majid Safari

[1]The University of Edinburgh, [2]University of Strathclyde, [3]University of Leeds, [4]University of Cambridge, [5]University of Bath, [6]Toshiba Europe Limited, [7]Nokia Bell Labs, [8]Oledcomm, [9]Du AUE, [10]Nav Wireless Technologies Pvt Ltd, [11]Crantec, [12]Orange, and [13]pureLiFi Limited

*This paper provides a summary of available technologies required for implementing indoor laser-based wireless networks capable of achieving aggregate data-rates of terabits per second as widely accepted as a sixth generation (6G) key performance indicator. The main focus of this paper is on the technologies supporting the near infrared region of the optical spectrum. The main challenges in the design of the transmitter and receiver systems and communication/networking schemes are identified and new insights are provided. This paper also covers the previous and recent standards as well as industrial applications for optical wireless communications (OWC) and LiFi.*

## Introduction

The number of wireless communication devices and their associated services are increasing rapidly requiring substantial improvements in data density measured in bits per second per square metre (bps/m$^2$). Key new applications will center on the cyber-physical continuum or metaverse including the convergence of sensing, communications, computing/artificial intelligence (AI) and robotics/control. It is generally accepted that in order to enable this vision, new spectrum beyond the classic radio frequency (RF) spectrum is required [1].

OWC is a viable solution to achieve the ambitious key performance indicators (KPIs) of 6G in terms of data density, peak data-rates, latency, energy efficiency and security. OWC enables the creation of ultra-small cells in a cellular network design. This is vital to unlock the path to exponential capacity growth for indoor scenarios and to achieve step-change improvements in data density. Furthermore, OWC can offload traffic from congested RF wireless networks, thereby providing extra capacity which can be traded off for a reduction in latency for time critical Internet of Things (IoT) applications. OWC will co-exist with current and future RF wireless communication systems to combat issues such as blockage (e.g., when an OWC-enabled smartphone is in user's pocket). Nevertheless, the availability of three orders of magnitude unregulated spectrum (close to 589 THz for VLC and IR) compared with the entire RF spectrum (close to 300 GHz) and using technologies that underpin optical fiber communication, OWC will significanty augment the RF communications. Because of the directionality of light and significant link quality degradations in a non-line-of-sight scenario, OWC has primarily been used for intentional communications in the past such as fixed point-to-point communications. IBM research in 1980 developed the first OWC networks between static computers using infrared (IR) light [2]. In the last two decades, however, OWC networks supporting user mobility have been developed where mostly visible light is used for both illumination and downlink communication while infrared light is used for uplink communication. The simultaneous function of illumination and downlink communication was motivated by the replacement of incandescent light bulbs with light emitting diode (LED) light fixtures. LEDs can achieve up to gigabit per second (Gbps) transmission speeds while user mobility is enabled by a careful integration of link diversity techniques, both at the access point and the mobile device. Mobile OWC networks are also referred to as LiFi networks [3].

A limitation of LEDs is their limited electrical bandwidth. In contrast, laser diodes achieve significantly higher bandwidth, but eye safety limits their output power. However, recent work has shown that vertical-cavity surface emitting laser (VCSEL) can be used to achieve tens of Gbps transmission speeds while maintaining eye-safety limits. When combining this capability with the directionality of light, it has been shown that it is possible to build LiFi access points that can achieve aggregate data rates of greater than 1 Tbps in line with 6G KPIs [4]. VCSELs exhibit advantageous features such as high-speed direct modulation (ten's of GHz bandwidth – a single VCSEL could have as much bandwidth as 1/6 of the entire RF spectrum), high power conversion efficiency, low cost, long lifetime and possibility to manufacture highly integrated VCSEL arrays. These attributes make VCSELs very attractive for use in many applications, particularly for high-speed indoor networks.

## State-of-the-art Technologies

### Transmitter technologies

This section reviews the different types of optical sources which have been used mainly for indoor OWC transmitters over the years. These sources operate over a wide spectrum ranging from visible (from 400 nm to 700 nm) to near infrared (NIR) (up to 1550 nm) wavelengths. LEDs and laser diodes (LDs) are the two main types of sources considered in OWC. LEDs, given their rising utilisation for indoor lighting, have been widely adopted in visible light communication (VLC) transmitters.

Phosphor converted LED (PC-LED), multi-chip or multi-primary LEDs, micro-LEDs ($\mu$-LED) and organic LEDs (OLED) are various kinds of LEDs that have been proposed for lighting. PC-LEDs are cheap with limited data-rate, $\sim 1$ Gbps, while multi-chip LEDs generally combine three or more sources emitting different colors, typically red, yellow, green and blue (RYGB). Multi-chip LEDs offer higher aggregate data-rates with wavelength division multiplexing (WDM). An advanced form of LEDs, i.e., the $\mu$-LEDs, can achieve modulation bandwidths in range of hundreds of MHz, and therefore provide high data-rates. The $\mu$-LEDs have a small size ($\leq 100$ $\mu$m), when compared to other types of LEDs and therefore, rely on a relatively large number of sources to provide sufficient illumination for indoor lighting purposes. OLEDs are a popular technology for flat panel displays, in which the light is generated from organic thin films sandwiched between an anode and a metal cathode. The typical modulation bandwidth of OLEDs is around hundreds of kHz. Lastly, quantum dot LED (QLED) with different materials and processes has shown very impressive results so far for bright display panels which can offer modulation bandwidth up to MHz, making them less capable as the optical transmitter for multi-Gbit/s communication.

In contrast, due to their higher bandwidth, LDs have been used as transmitters in high-speed VLC, infrared OWC and fiber-optic communication systems. The LD output is typically more coherent and better collimated than LEDs, which is useful for improved efficiency and point-to-point data transmission. Another unique advantage of LDs over LEDs is the narrow optical emission spectrum which can allow high aggregate data-rates in conjunction with WDM.

### Receiver technologies

Photodetectors (PDs) are semiconductor devices that convert light to electrical current and are used as the main part of OWC receivers along with optical elements such as a lens. Various types of PDs are available including silicon (Si), germanium (Ge), gallium–arsenide (GaAs), gallium–aluminium–arsenide (GaAlAs), and Indium gallium arsenide (InGaAs). Each of these detector technologies have a unique spectral responsivity and quantum efficiency characteristics when operating in different wavelengths. It is typical for PDs that the responsivity exhibits a strong wavelength dependence.

Silicon PDs, for instance, have their highest sensitivity in the NIR region of optical spectrum between 800 nm and 1000 nm, with a relatively steep decline of the responsivity for higher wavelengths and very little responsivity beyond 1100 nm.

Silicon PDs are cheaper than other PDs as Silicon is largely found in nature in the form of sand. In contrast, Ge is not so easily found in nature, and when it is, it is only found in chemical compounds. Silicon PDs have the widest use among different technologies of PDs. Silicon PDs are sensitive in the near ultraviolet (UV), visible and NIR up to about 1100 nm; for higher wavelength ranges, other technologies, such as InGaAs and Ge PDs are in use. The Ge PDs have advantages over Si PDs in terms of energy loss and current loss, where the Ge PDs loss is lower than (almost half of) Si PDs.

Silicon PDs have a larger band-gap compared to Ge PDs, therefore, the phenomenon of thermal pair generation is smaller in Si PDs than in Ge PDs. That is, at the same temperature, the noise of the Si PDs is smaller than the noise of Ge PDs. Also, the reverse current of p-n junction is smaller for Si than for Ge. The noise of GaAs is smaller than Si and Ge. Table 1 represents the band-gap values for different semiconductors in which higher values of band gap correspond to lower noise. Compared to other PDs, GaN has a wider band gap and they are a good choice at UV frequencies. They can operate at much higher temperatures and work at much higher voltages than GaAs transistors [5].

| PD | Band gap [eV] |
|---|---|
| GaN | 3.4 |
| GaAlAs | 1.42-2.16 |
| GaAs | 1.43 |
| Si | 1.12 |
| Ge | 0.67 |
| InGaAs | 0.36-1.43 |

**Table 1** *Band gaps for different types of semiconductors (higher band gap means lower noise).*

## Laser-Based Communications

### Laser safety considerations

IEC 60825-1 is applicable to the safety of laser products emitting radiations in the wavelength range from 180 nm to 1 mm. For the infrared range where a high availability of off-the-shelf laser sources exist, the major cause of damage is thermal. For mid and far infrared wavelengths i.e., longer than 1400 nm, the main consideration is to avoid damaging the cornea and the skin, as radiation at those wavelengths is absorbed by water, and the vitreous humor of the eye protects the retina from damage [6].

In the NIR (700 nm – 1400 nm), although care must be taken to avoid thermal damage to the skin, the predominant consideration is to avoid retinal damage. The ability of the cornea and the lens to focus these wavelengths directly onto a small spot on the retina means that these wavelengths can be significantly more hazardous than the longer infrared wavelengths, even at lower optical irradiance at the surface of the eye. There are three additional factors other than wavelength to consider when evaluating the safety of an infrared laser-based optical wireless communications system: the duration of exposure, the most hazardous position, and the angular subtense of the apparent source, which is defined as the angle subtended by the apparent source as viewed from a point in space.

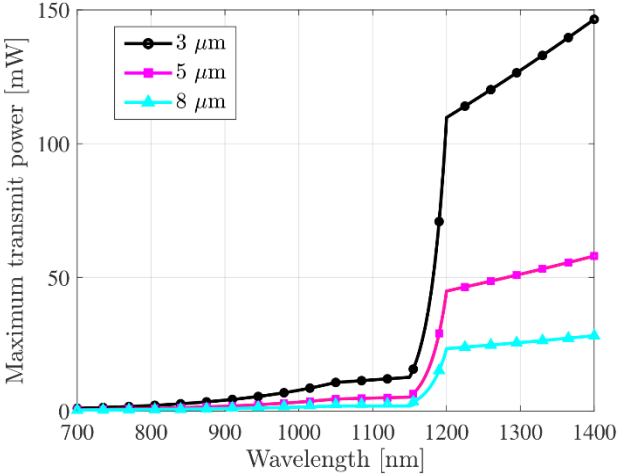

**Figure 1** *The maximum transmit power versus wavelength assuming long exposure durations.*

Figure 1 presents the maximum transmit power that satisfies eye safety regulations of IEC 60825-1 standard versus the wavelength of the laser source for long exposure durations. Results are provided for three practical ranges of the laser beam waist. The results show that as the operating wavelength of a laser source increases, the maximum transmit power also increases. In particular, the maximum transmit power increases exponentially beyond wavelength 1050 nm and up to wavelength 1200 nm, where the eye safety constraint is much more relaxed in comparison to lower wavelengths. For wavelengths higher than 1200 nm, the maximum transmit power increases linearly as the wavelength increases. Another observation is the relationship between the beam waist and the maximum transmit power. It can be inferred that the maximum transmit power reduces as the beam waist increases, which is due to the fact that the laser beam is less divergent for a larger beam waist.

## Communication System Design and Link Budget Considerations

Depending on the beam divergence, the OWC link is generally categorised as narrow-beam or broad-beam. The capabilities of these two categories are different and therefore, each category addresses different set of applications using different types of sources, detectors and optics. Narrow-beam applications focus on very high data-rate (typically higher than 10 Gbit/s) fixed and nomadic links for applications such as data-centers, indoor wireless access and free-space optical links. The beam divergence in such links is in the order of ∼ 0.1°. Broad-beam OWCs are typically used for short-range wireless multi-user access requiring ∼Gbit/s rates. Provided a receiver with wide field-of-view (FOV) is available, the broad-beam OWC links can support applications requiring mobile user/machine without the need of tracking and beam-steering. Therefore, the broad-beam links can be more compact and cost-effective than their narrow-beam counterparts. In the following, we present two scenarios including a point-to-point fixed backhaul OWC link and an OWC access link providing coverage for an extended indoor space.

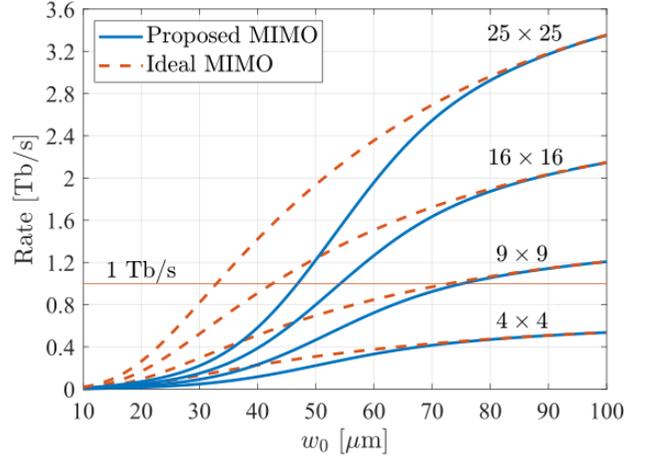

**Figure 2** *Aggregate data-rate of the proposed MIMO optical wireless backhaul system as a function of beam waist radius, $w_0$.*

**1) Backhaul OWC Scenario:**

In a 6G network, the backhaul system is required to support aggregate data-rates of Tb/s generated by the access network. To this end, a multiple input multiple output (MIMO) OWC backhaul link can be considered by using a VCSEL array at the transmitter and a PD array at the receiver. Figure 2 illustrates the aggregate data rate achieved by the proposed MIMO system over a link distance of 2 m when the radius of the beam waist, $w_0$, varies between 10 µm and 100 µm. The parameters are taken from [7]. For comparison, an ideal MIMO system is used, which consists of parallel single input single output (SISO) channels with no crosstalk among them. This baseline scenario provides the upper bound performance of the MIMO system. For all MIMO realizations shown in Figure 2, the aggregate rate monotonically increases with $w_0$. Note that in far field, the radius of the beam spot is inversely proportional to beam waist. Thus, the right tail of the curves indicates the noise-limited region for $w_0 \geq 80$ µm, whereas the left tail for $w_0 < 80$ µm represents the crosstalk-limited region. For $w_0 = 10$ µm, the beam spot size at the receiver is very large, i.e., 54 mm. Such a large beam spot renders the signal power collected by each PD from direct channels very low and causes substantial crosstalk among the incident beams. By increasing $w_0$, the gap between the performance of the MIMO system asymptotically approaches the upper bound. To achieve a target data rate of higher than 1 Tb/s, 9 × 9, 16 × 16 and 25 × 25 MIMO systems require beam waist radii of at least 75 $\mu$m, 54 $\mu$m and 47 $\mu$m, respectively. These effective beam waists can be realized by placing a convex lens in front of each VCSEL with a proper optical design. The 1Tb/s target is not achievable by a 4 × 4 system for $w_0 \leq 100$ $\mu$m.

### 2) Multi-User Access Scenario:

For indoor multi-user wireless access networks, laser-based OWC systems with beam steering can achieve multi-Gb/s data-rates over the confined spot of each beam by aiming the laser beam at the user's device. These systems apply advanced optical beam steering designs to provide IR laser beam-class indoor coverage. However, the high complexity and expensive components in such designs may not suit large-scale deployments for indoor networks. Alternatively, a novel access point (AP) architecture without the need for beam steering was proposed in [8] based on an array of VCSEL arrays.

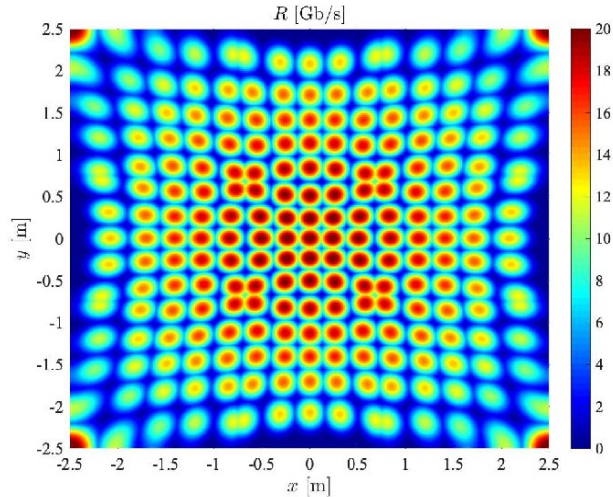

**Figure 3** *Spatial distribution of the data-rate on the receiver plane at 3 m distance from the AP.*

With the considered array of arrays architecture, data-rates of at least 10 Gb/s per beam can be realized using direct current optical orthogonal frequency division multiplexing (DCO-OFDM), leading to aggregate data-rates greater than 2 Tb/s from a single AP. The system performance is evaluated in a 5 × 5 × 3 m³ room using OpticStudio simulations. The AP with 9 VCSEL arrays each consisting of 25 VCSELs (i.e., 5 × 5 arrays), is mounted at the center of the ceiling to cover the floor area of 25 m². Each VCSEL array has a total area of 1 cm² and it is responsible for covering almost 1/9 of the room area. To separate the beam spots on the receiver plane, a plano-convex lens is placed in front of each array to refract the output lights from different VCSELs towards different directions. The desired refraction angles for each VCSEL can be ensured by properly adjusting the lens parameters. A beam waist radius of 5 $\mu$m, a transmit power of 10 mW per VCSEL, a laser wavelength of 950 nm, a modulation bandwidth of 5 GHz, and an effective PD area of 2 cm² are assumed in these simulations. Also, we assume that the receiver plane is located 3 m below the AP. Figure 3 demonstrates the spatial distribution of data rates, where data rates of 15 to 20 Gb/s are achieved at the beam spot centers. Besides, a data-rate of at least 10 Gb/s is guaranteed almost everywhere on the receiver plane, except at the beam spot borders and in locations very close to the walls. Providing at least 10 Gb/s per beam, the AP delivers the aggregate data-rate beyond 225 × 10 Gb/s = 2.25 Tb/s. Note that coverage at beam borders can be realized through proper interference management schemes.

## Networking Architecture for Laser-Based LiFi

In this section, the advanced techniques relevant to the architecture of indoor laser-based wireless communication networks are discussed, addressing issues such as cell formation, interference management, resource allocation, and backhauling.

### Cell formation

In laser-based networks with very small cells, users with high mobility may be subject to frequent handovers, which then results in the need for resilient handover schemes ensuring continuous quality of service (QoS). Network-centric (NC) approaches such as in [9] can address the limitations mentioned above in terms of the small coverage area of optical APs and enhancing signal-to-noise ratio (SNR). In this design, the whole receiving plane is divided into multiple static cells, each with multiple APs that jointly serve multiple users. In contrast, a user centric (UC) network topology was designed in [10] to avoid the inter-cell interference (ICI) among multiple optical cells and achieve high data-rates. The UC cells are characterized by their elastic shapes where they are formed as a function of the distribution of users. Therefore, the UC design is updated over time in responding to any changes into the network resulted from the mobility of users. In comparison to the NC design, the UC approach involves high complexity comprising additional signaling. However, it is worth implementing since it enhances the performance of optical networks significantly in terms of SNR and achievable data rates.

### Interference management

In laser-based cellular networks, serving a high number of users requires efficient interference management schemes to align data streams transmitted to users with a minimum error. Orthogonal transmission schemes such as time division multiple access (TDMA), space-division multiple access (SDMA) and orthogonal frequency division multiple access (OFDMA) were implemented to manage the interference among multiple users through transmitting data streams in an orthogonal fashion giving users the ability to decode their information independently in the absence of interference in their allocated time, frequency or space channels. Despite the low complexity of these schemes, they may not efficiently exploit the capacity of laser-based cellular networks due to the fact that users cannot be served using the same frequency or time slots. Another orthogonal approach that is particularly suited for OWC technology is WDM, which has been used in [11] by employing RGBY LEDs, by allocating different colors to different users. In this scenario, optical filters must be used on the receiver side to differentiate various wavelengths. However, the broad emission spectrum of LEDs can result in crosstalk among different wavelengths even in the presence of optical filters. This issue will not affect the performance of multi-user laser-based communication to the same degree due to lasers narrower emission spectrum.

*Resource allocation*

Given the remarkable developments and rapid evolution of communication networks and computing architectures, resource allocation continues to be a non-trivial matter. The developed resource allocation framework must be able to cope with various new emerging architectures such as wired/wireless mesh networks, multi-protocol networks, cognitive networks, and cloud/fog computing systems. Over the recent years, various optimization techniques have been utilized for efficient resource allocation including centralized solutions such as mixed integer linear programming and distributed ones such as machine learning, game theory, and auction theory, to name a few. Therefore, given the complex nature of communication systems, it becomes imperative to design advanced resource allocation methodologies to ensure users required QoS.

*Backhaul Solutions*

Besides the OWC backhaul solution mentioned in the last section, a natural option for realizing high-capacity backhaul links is the distribution of optical fiber cables (e.g., Single Mode Fibers, Multimode fiber or Plastic Optical Fiber) in the indoor environment where the Tb/s access network capacity is required. In a star topology, every AP is individually equipped with an optical fiber link that connects it to the fiber distribution hub through which it is routed to the Internet. However, this is feasible if the expenditure incurred due to redesigning the wiring infrastructure is justifiable (for new buildings, or where it is easy to deploy fibers in an apartment). Around 80% of Internet traffic was reported to initiate and terminate indoors. Thus, it would improve such a fiber distributions network if its connections are in the form of a data network (i.e. replacing the star topology with different network topologies like a tree tropology or a ring topology). For example a ring network topology can then meet these indoor demands which can be in the form of user-to-user and IoT-to-IoT communication or Machine to machine communication in addition to user-to-Internet and vice-versa traffic through the introduction of broadcast and select mechanisms.

In WDM passive optical network (PON) systems, each subscriber has a physical interface corresponding to its guaranteed bandwidth. For physical tree topologies, the aggregation is made using optical multiplexers or optical couplers to transport the data of several subscriber onto a same fiber. This solution can require tunable lasers, when optical couplers are used, to adapt the wavelength, in particular if the transport fiber is shared with another PON system.

The optical solutions offer low power consumptions when compared to other technologies for a deployment in highly energy sensitive environments. This is due to the fact that technologies such as active Ethernet architecture require several distributed switches (workgroup and core) and therefore consume more power. Moreover, Figure 4 shows that the total cost of ownership for a passive optical Local area network (LAN) is only 43.7% of that for a traditional LAN with the same parameters.

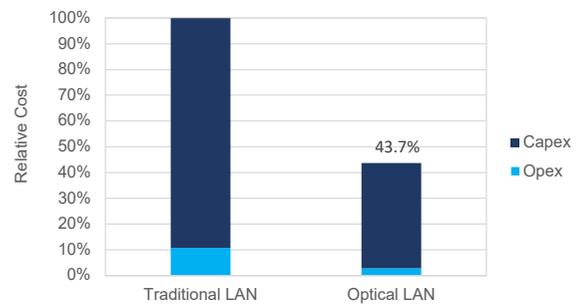

**Figure 4** *Total cost of ownership in terms of capital expenditure (Capex) and operational expenditure (Opex) for a traditional vs a Nokia passive optical LAN.*

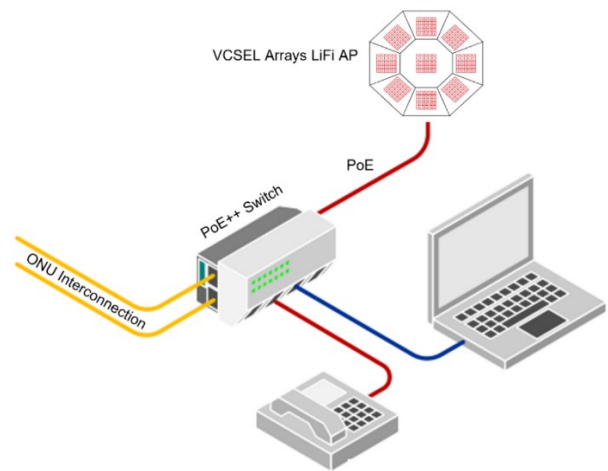

**Figure 5** *Indoor PoE++ LiFi wireless LAN network connecting a VCSEL arrays LiFi AP.*

As a cost-effective solution, the IEEE 802.3bt PoE++ is becoming the standard for connected homes and buildings. Proposing a network topology where LiFi APs are connected using a PoE++ Gigabit Ethernet will reduce the cost of integration, deployment and maintenance. We suggest a new networking option, which consists of an interconnecting optical network unit (ONU) with a PoE++ switch equipped with the small form-factor pluggable (SFP) ports. This interconnection will then create a 802.3bt PoE++ Gigabit Ethernet local area network (LAN) backbone to deploy a LiFi wireless network as illustrated in Figure 5. In a distributed architecture, LiFi APs will be connected to the 802.3bt PoE++ network using the industrial micro switch. If VCSEL array LiFi APs are then equipped with a PoE network connectivity it will be easy and low cost to connect it to a Gigabit Ethernet switch as shown in Figure 5.

## Current Standards and Industrial adaptations

Initial attempts to standardize commercial OWC technologies started in the 1990s. Infrared data association (IrDA) was formed in 1993 to develop interoperable IR-based OWC. This was then followed by the formation of IEEE 802.11 in 1997, which can achieve a maximum speed of 2 Mbps. This legacy standard also covers specifications of data transmission over the IR spectrum. In 1999, due to its limited ability to provide high data rates and wide coverage, IR communications were quickly superseded by IEEE 802.11b (also known as WiFi 1), which could achieve a maximum speed of 11 Mbps. In 2000, research on OWC shifted to focus on the visible light spectrum due to the invention of highly efficient blue LEDs [12]. This new interest led to standards in VLCs, e.g., the VLC Consortium, IEEE 802.15.7, IEEE 802.15.13 and International Telecommunication Union - Telecommunication Standardization Sector (ITU-T) G.9991 (also known as G.vlc).

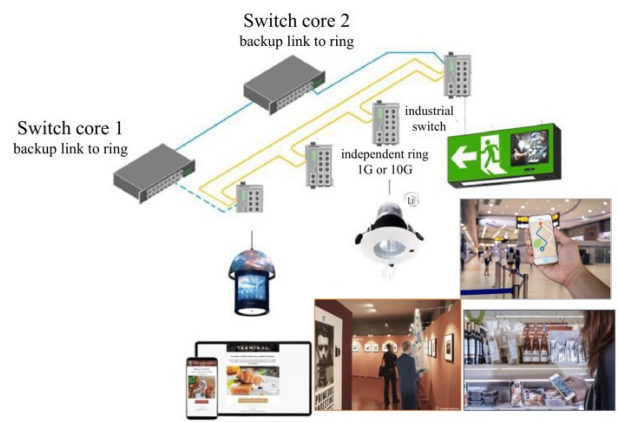

**Figure 6** *Indoor PoE++ LiFi wireless LAN network mobile applications & services.*

The Japan electronics and information technology industries association (JEITA) published two VLC standards, namely the JEITA CP-1221 and JEITA CP-1222 in 2007. Since 2009 on-wards, the standardization efforts were mainly concentrated within the IEEE and more precisely within its 802.15 working group, first through the task group (TG) 7 [13]. In 2011, the IEEE 802.15.7-2011 standard inspired by ZigBee was published, where the TG7 initiates a revision to this standard draft in 2015. In January 2015, IEEE 802.15 formed a task group to write a revision to IEEE 802.15.7-2011 that accommodates infrared and near ultraviolet wavelengths, in addition to visible light, and adds options such as optical camera communications (OCCs) and LiFi.

OCCs enable scalable data-rate, positioning/localization, and message broadcasting, etc. using devices such as the flash, display and image sensor as the transmitting and receiving devices. OCC can be used in museums and retailers, where OCC offer less expensive audio guides as well as being a new way to share lighting costs with brands and as an opportunity to offer promotional content and coupons to customers (Figure 6). In November 2017, the IEEE created a new study group to drive worldwide standardization work on light communication standards and to introduce LiFi to the market. The LiFi standardization journey started since then has seen an increasing interest and a steady growth of new stakeholders. In July 2018, a new IEEE task group focusing on light communications was formed, namely IEEE 802.11bb (TGbb). The primary motivation of this task group is to standardize mobile, networked light communications, i.e., LiFi.

Today, the main applications for infrared wireless communication may be divided into five categories based on the transmission range:

- Ultra-short range: chip-to-chip communications in stacked and closely packed multi-chip packages.
- Short range: wireless body area network (WBAN) and wireless personal area network (WPAN) applications under standard IEEE 802.15.7.
- Medium range: indoor IR for wireless local area networks (WLANs) and inter-vehicular and vehicle-to-infrastructure communications.
- Long range: free-space optical communications (FSO) and underwater communications. In comparison to its acoustic and RF counterparts, underwater optical wireless communication can provide a much higher transmission bandwidth and as a result much higher data rates [14].
- Ultra-long range: laser communication in space especially for inter-satellite links and establishment of satellite constellations. This application will be a key enabler for 6G space-air-ground integrated networks.

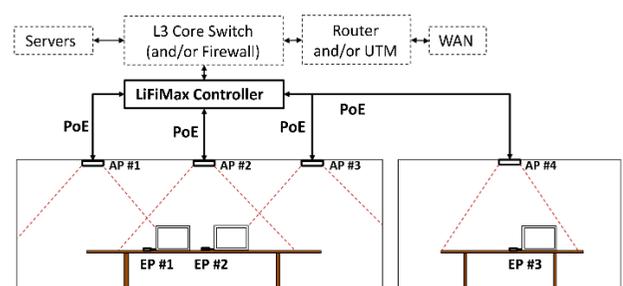

**Figure 7** *The LiFiMAX solution connection diagram.*

Industrial companies have developed products and prototypes for several different use cases, from indoor networking to space applications. For example, the LiFiMAX product has been introduced, based on the ITU-T G.vlc standard. The LiFiMAX solution is shown in Figure 7, which has been deployed in more than five hundred projects to enable indoor access networks. As represented in Figure 7, each AP installed on the ceiling generates a fixed cell in which any end point (EP, i.e. device equipped with a compatible dongle) can connect to the corresponding AP and thus to the network. The LiFiMAX solution also supports mobile EP thanks to embedded handover management mechanisms [13]. In parallel, industry has participated in research works on the use of light communication for extra-WBAN applications, with a focus on power efficient multiple access schemes which are able to fulfil the requirements of such a use case and their experimental evaluation [15].

## Conclusions and Future Research Directions

The available technologies suitable for the design of an indoor laser-based wireless network capable of achieving users' QoS significantly beyond the 5G requirements have been studied. Transmitter and receiver technologies across different wavelengths of infrared light were compared and laser safety limitations were summarized. Laser safety requirements provide different link budget margins for achieving reliable connectivity in an indoor environment. For example, 1550 nm technology can achieve higher data-rates compared to the 800 nm $-$ 1000 nm technology due to more relaxed laser safety requirements. However, the attractive features of VCSEL-based sources such as high-speed direct modulation, low cost, energy efficiency, and ease of manufacturing as arrays, bring the 800 nm $-$ 1000 nm technology at which most VCSEL sources operate, into the focus of research. These features are not only attractive from a practical and commercial perspective, but also provide some exciting opportunities to design novel communication and networking schemes based on a multiple-narrow-laser-beam network architecture.

Future research directions aim to tackle the main challenges involved in the design of a terabit indoor laser-based wireless network including:
1) Effectively exploiting the advantages of VCSEL-based sources in the form of arrays, using proper optical design and electrical adaptive control,
2) Designing optical receivers based on detector arrays with high sensitivity and sufficient FOV to support user mobility,
3) Developing novel modulation and signal processing schemes that can provide reliable data rates beyond 10 Gbps in a laser-safe power-limited wireless network,
4) Developing networking schemes that can provide seamless connectivity of mobile users in an indoor laser-based narrow-beam network supporting an aggregate throughput of terabit per second.

## Acknowledgements

The authors acknowledge financial support from the EPSRC under program grant EP/S016570/1 'Terabit Bidirectional Multi-User Optical Wireless System (TOWS) for 6G LiFi'.